\documentclass[12pt]{article}
\usepackage{amsmath}
\usepackage{graphicx,psfrag,epsf}
\usepackage{enumerate}
\usepackage[numbers]{natbib}
\makeatletter 
\renewcommand\@biblabel[1]{#1.} 
\makeatother

\newcommand{\blind}{1}

\usepackage{macro}

\usepackage[english]{babel}
\usepackage{nameref, hyperref} 
\usepackage{graphicx, subfigure, caption} 
\usepackage{adjustbox}
\usepackage{threeparttable}
\usepackage[T1]{fontenc}
\usepackage[utf8]{inputenc}
\usepackage{filecontents}
\usepackage{authblk}
\usepackage{epsfig}
\usepackage{bmpsize}
\usepackage{rotating}
\usepackage{bibentry}

\usepackage{centernot}
\usepackage{booktabs}
\usepackage{lipsum}
\usepackage{algorithm, algorithmicx}
\usepackage{algpseudocode}
\usepackage{color}
\usepackage{amsfonts, amsthm, bm, amssymb}
\usepackage{caption}
\usepackage{cleveref}
\usepackage{multirow}
\usepackage{xr}
\begin{document}

\def\spacingset#1{\renewcommand{\baselinestretch}%
{#1}\small\normalsize} \spacingset{1}

\if1\blind
{
  \title{\bf A Mixture Model to Detect Edges in Sparse Co-expression Graphs}
  \author{
  	Haim Bar\thanks{haim.bar@uconn.edu}\hspace{.2cm}\\
    Department of Statistics, University of Connecticut\\
    and \\
    Seojin Bang\thanks{seojinb@cs.cmu.edu}\hspace{.2cm}\\
    Computational Biology Department, Carnegie Mellon University}
\date{}

  \maketitle
} \fi

\if0\blind
{
  \bigskip
  \bigskip
  \bigskip
  \begin{center}
    {\Large \bf A Mixture Model to Detect Edges in Sparse Co-expression Graphs}
\end{center}
  \medskip
} \fi


\begin{abstract} 
In the early days of gene expression data, researchers have focused on gene-level analysis, and particularly on finding differentially expressed genes. This usually involved making a simplifying assumption that genes are independent, which made likelihood derivations feasible and allowed for relatively simple implementations. In recent years, the scope has expanded to include pathway and `gene set' analysis in an attempt to understand the relationships between genes. We develop a method to recover a gene network's structure from co-expression data, measured in terms of normalized Pearson's correlation coefficients between gene pairs. We treat these co-expression measurements as weights in the complete graph in which nodes correspond to genes. To decide which edges exist in the gene network, we fit a three-component mixture model such that the observed weights of `null edges' follow a normal distribution with mean 0, and the non-null edges follow a mixture of two lognormal distributions, one for positively- and one for negatively-correlated pairs. We show that this so-called $L_2N$ mixture model outperforms other methods in terms of power to detect edges, and it allows to control the false discovery rate. Importantly, the method makes no assumptions about the true network structure.
\end{abstract}

\noindent%
\vfill

\newpage
\setcounter{page}{1}
\spacingset{1.45} 


\section{Introduction}\label{sec.introduction}

Broadly speaking, statistical analysis of `omics' data consists of studying
the relative abundance of biological `building blocks', such as genes, proteins, 
and metabolites. The goal of many studies involving high-throughput data is to
identify differential building blocks - those whose abundance levels vary 
according to the value of some other factor.
These factors include, for example, environmental conditions, disease state, 
gender, or age.
To simplify the discussion, we will 
use genomics terminology, where 
the building blocks are genes and the abundance is their expression level.
Many biological studies use statistical methods which focus on individual genes 
and rely on the unrealistic, but mathematically convenient assumption that the 
expression levels are independent across genes. 
However, other methods drop this assumption and acknowledge that multiple genes are 
likely to work as a group associated with the same biological process, thus providing
not only more complex functionality, but also robustness to detrimental 
mutations.
A common assumption 
 is that related genes share a 
regulatory process, and therefore, their expression levels are expected to be highly 
correlated.
The relevant measurements in this context are \textit{co-expression} 
levels of pairs of genes, which can be measured in terms of Pearson's correlation coefficient, 
Mutual Information, Spearman's rank correlation coefficient, or Euclidean distance.
Gene co-expression data can be seen as a network; an undirected complete graph 
in which nodes represent genes and weights are assigned to edges according to the 
strength of the association between each pair's expression levels, across multiple samples. 

Using gene co-expression patterns, a number of authors defined `modules' as sets of genes
that have similar expression patterns; they then focused on a small number of intramodular `eigengenes' or `hub genes' instead of on thousands of genes \citep{stuart2003gene,zhang2005general,eisen1998cluster,taylor2009dynamic,barabasi2011network}. Among them, the weighted gene co-expression network analysis (WGCNA, 
\citep{zhang2005general}) is a widely used approach. Among hub genes, the aforementioned independence assumption is more reasonable because genes belonging to different modules are expected to be much less correlated than genes within the same module. Thus, one can try to find differentially expressed hub genes with respect to a trait or treatment. 
However, methods assuming a specific structure and focusing on modules and their eigengenes or hub genes, are only suitable for certain networks where nodes within a module are highly connected but connections across modules are relatively rare. However, biological networks such as protein-protein and gene-gene interaction networks may have other network features where modules cannot be clearly partitioned. Such is the case, for example, with scale-free networks \citep{ravasz2002hierarchical, wuchty2003evolutionary, tong2004global}, a scale-free regime followed by a sharp cutoff  \citep{newman2001scientific, amaral2000classes, jeong2001lethality}, and other networks with curved degree distributions \citep{zhang2011network, chu2012constructing, smith2006network, radrich2010integration}.
Instead of identifying hub genes and comparing their differential expression levels between two traits or treatments, there have been other approaches that compare high dimensional covariance matrices between the two groups and identify risk genes based on the correlation structures \citep{schott2007test,li2012two,cai2013two,cai2016large,zhu2017testing}.
Recently, \citet{cai2016large} suggested a method for large-scale testing of correlations under certain regularity conditions. \citet{zhu2017testing} suggested a sparse leading eigenvalue driven test that compares two high-dimensional covariance matrices obtained from schizophrenia and normal groups and identified novel schizophrenia risk genes. Such genes that are found to be highly associated with a trait or treatment are further analyzed by using knowledge-based pathway analysis tools such as Gene Set Enrichment Analysis (GSEA, \cite{subramanian2005gene, mootha2003pgc}). Pathway analysis differs from co-expression analysis in that it uses \textit{pre-defined} gene sets 
(e.g., from the Gene Ontology (GO, \cite{gene2004gene}) or the Kyoto Encyclopedia of Genes and Genomes (KEGG, \cite{kanehisa2000kegg}). Such analyses help determine which pathways are over- or under-represented in the identified modules \citep{khatri2012ten}.

Obtaining accurate estimates of covariance (or precision) matrix is important for gene network analyses. However, it becomes challenging when the number of genes is larger than the sample size (as is often the case). The problem of estimating large, sparse gene networks has been thoroughly studied in modern multivariate analysis. Researchers have suggested various approaches using regularization techniques, and one of most commonly used approaches is the penalized maximum likelihood. For example, \citet{meinshausen2006high} proposed to estimate a precision matrix by imposing an $l_1$ penalty on a Gaussian log-likelihood to increase its sparsity. It uses a simple algorithm to estimate a sparse precision matrix, by fitting a regression model to each variable with all other variables as predictors, and apply the lasso to obtain sparsity. \citet{friedman2008sparse} suggested a simple but faster algorithm called graphical lasso, which also estimates a sparse precision matrix using coordinate descent procedure for lasso. \citet{yuan2007model, banerjee2008model, rothman2008sparse} and \citet{levina2008sparse} also proposed algorithms to solve the $l_1$ penalized Gaussian log-likelihood to estimate a large, sparse precision matrix. These methods have been extensively used to estimate sparse gene network. 
The use of precision matrix for estimating gene network is justified when the data is generated from a multivariate normal distribution. Under this assumption, elements of the precision matrix correspond to conditional independence restrictions between nodes in a gene network. These approaches work well when analyzing gene data obtained from large retrospective studies such as the TCGA data \citep{TCGA}. However, applying these approaches to datasets with a relatively small number 
of samples may not yield robust estimates of the regression coefficients. 

The main goal of this paper is to introduce a new method to uncover the structure of sparse gene networks from expression data. 
Rather than estimating the precision matrix, we use the covariance matrix to determine which genes are co-expressed. 
Sparsity in the covariance matrix (rather than the precision matrix) is motivated mostly by biological arguments -- (i) expression level of genes are associated with their functionality, and thus, are expected to be highly correlated for genes that share the same functional process; and (ii) the number of genes associated with most biological functions is small, relative to the total number of genes.
There is also a strong mathematical argument for modeling sparsity in the covariance matrix, rather than the precision matrix. 
For each gene, its expression profile is an $N$-dimensional vector where $N$ is the sample size, and for each pair of genes the correlation between their expression profiles represents the cosine of the angle between the two corresponding vectors. So, the correlation matrix offers an intuitive representation of the similarity between expression profiles.
As \citet{Frankl1990} showed, as $N$ increases, any two random vectors in the Euclidean $N$-dimensional space are approximately orthogonal with probability that approaches 1. Thus, the correlation between two random expression profiles is expected to be close to $cos(\pi/2)=0$, which means that the correlation (and hence, the covariance) matrix is expected to be sparse.


While gene expression datasets consist of thousands of genes and the complete network contains millions of edges, in most biological systems gene networks are expected to be sparse  \citep{yeung2002reverse}.
With a finite sample, however, the observed correlations are not zero, even for uncorrelated
pairs, so the first problem is to define an accurate decision rule to
differentiate between spurious correlations and true edges.
A second, related problem, is how to perform the computation needed to implement
such a rule in practice.

To address these challenges, we first obtain weights, $w_{ij}$, by applying 
Fisher's Z transformation to the sample correlation coefficients, $r_{ij}$. For uncorrelated pairs, the asymptotic 
distribution of $w_{ij}$ is normal, with mean zero and variance $N-3$.
This motivates fitting a mixture model to $\{w_{ij}\}$ in which the majority of pairs belong to a
normally distributed `null component', and a small percentage of the weights belong to one of two `non-null
components', which follow log-normal distributions (one for positive and one for negative
correlations). This so-called $L_2N$ model was first presented in \citet{BarSchifano}, in the context
of identifying differentially expressed (or dispersed) genes. We use the $L_2N$ model to recover a sparse gene network for the following reasons. 
First, the $L_2N$ mixture model leads to shrinkage estimation and to borrowing strength across all pairs, 
which increases the power to detect co-expressed pairs. Second, the specific form of the mixture 
model allows us to establish a decision rule which controls the error rate. 
Third, the mixture model lends itself to a computationally-efficient estimation of the parameters via the EM algorithm \cite{EM}.
Note that when \citet{BarSchifano} used the model, they assumed independence between genes, while
this is not the case in our application. Furthermore, they used the $L_2N$ model in the k-group comparison setting,
while in our application we fit the model to detect a network for each group.

This paper is organized as follows: in Section \ref{sec.model} we present our method to uncover the structure of gene networks.
We discuss some computational challenges and our proposed solutions.
In Section \ref{simulations} we evaluate the goodness of fit of our method and its ability to correctly recover the network structure and compare our method's ability to correctly identify true edges with existing methods. 
In Section \ref{sec:casestudy} we apply our method to a publicly available collection of breast cancer datasets from twelve studies. 
We conclude with a discussion in Section
\ref{discussion}.  

\section{Statistical Model and Estimation}\label{sec.model}
\subsection{A Mixture Model for Edge Indicators in a Gene Network}\label{subsec.mixture}
A gene network can be represented by a weighted, undirected graph in which each 
node corresponds to a gene and each edge corresponds to a pair of genes that are 
`co-expressed', meaning that their expression levels are highly correlated.
The weights represent the strength of the connection between two genes, namely, their
tendency to be co-expressed.
Given normalized expression data of $G$ genes, our objective is to discover  the network structure,
namely, which of the $K=G(G-1)/2$ pairs are co-expressed. 
Our general strategy 
is to associate with every putative 
edge in the complete graph with $G$ nodes, a latent indicator variable whose value (0 or 1) 
is determined by a statistical model.

\begin{figure*}
\centerline{\includegraphics[scale=0.7]{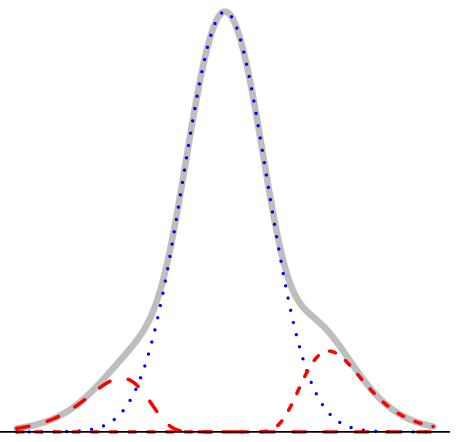}}
\caption{The $L_2N$ mixture model, with probability of the null component (blue dotted curve),
 $Pr(mn\in C_0)=0.8$, where $mn$ denotes the edge between nodes $m$ and $n$ 
in the graph.}\label{fig:mixture}
\end{figure*}

To start, we define the edge weights in terms of pairwise correlation coefficients. 
Let $\mathbf{x}_{g}$ be a vector of normalized expression levels for 
gene $g\in\{1,\ldots,G\}$ obtained from $N$
samples ($N>3$). Suppose that the true correlation coefficient between genes $m$ and $n$
is $\rho_{mn}$,
and let $r_{mn}=corr(\mathbf{x}_{m}, \mathbf{x}_{n})$ be the observed correlation coefficient.
Using Fisher's Z transformation, we obtain the estimated weight $w_{mn}=arctanh(r_{mn})$, which is 
known to be approximately normally distributed, with mean $arctanh(\rho_{mn})$ and variance 
$\frac{1}{N-3}$.
Let $E=\{e_{mn}\}$ be the set of true edges in the network.
We assume that $G$ is large and that most pairs are not co-expressed, 
so the network is sparse: $|E| \ll K$. 
This assumption, along
with asymptotic normality of $w_{mn}$, motivate our model choice. Specifically, we
assume that the weights follow the so-called $L_2N$ mixture distribution in \citet{BarSchifano}.
$L_2N$ is a three-component mixture model in which the `null' component follows 
a normal distribution with mean 0, representing the majority of 
pairs that have approximately zero correlation,
and the tails (the `non-null' components, for pairs with strong positive/negative correlations)
follow log-normal distributions:
\begin{eqnarray}
w_{mn} | e_{mn} \notin E &\sim& N(0, \sigma^2)\\
w_{mn}|[w_{mn}>0,\, e_{mn}\in E]\, &\sim& 
    LogNormal(\theta_1,\kappa_1^2)\label{L2NDE1}\,,\\\vspace*{-10pt}
-w_{mn}|[w_{mn}<0,\, e_{mn}\in E]\, &\sim& LogNormal(\theta_2,\kappa_2^2)\label{L2NDE2}\,.
\vspace*{-10pt}
\end{eqnarray}
Note that $\sigma^2$ consists of two variance components, namely, $\sigma^2=\frac{1}{N-3}+\sigma_0^2$, 
where $\frac{1}{N-3}$ is the variance component due to the asymptotic distribution of $arctanh(r_{mn})$, 
whereas $\sigma_0^2$ is due to the random effect model,
which allows us to account for extra variability among uncorrelated pairs.
A graphical representation of the $L_2N$ model is shown in Figure \ref{fig:mixture}.

If we denote the null mixture component in the $L_2N$ model 
by $C_0$, the two non-null components by $C_1$ and $C_2$,
 the corresponding probability density functions by $f_j$, and the mixture probabilities by $p_j$,
 for  $j=0,1,2$, such that $p_0+p_1+p_2=1,$ then we classify a 
putative edge between nodes $m$ and $n$ in the complete graph into one of the three mixture 
components based on the posterior probabilities,
\begin{equation}\label{eq:posterior_prob}
Pr(e_{mn}\in C_j | w_{mn}) = \frac{p_jf_j(w_{mn})}{p_0f_0(w_{mn})+p_1f_1(w_{mn})+p_2f_2(w_{mn})},~j=0,1,2.
\end{equation}

Let $\mathbf{b}_{mn}=(b_{0mn},b_{1mn},b_{2mn})$ be an indicator vector, so that $b_{jmn}=1$ for the
component $j$ with the highest probability, $Pr(e_{mn}\in C_j | w_{mn})$, for the pair $mn$, and 0 for the other two components. Using this notation, the $G\times G$ matrix
$\textbf{A}=[1-b_{0mn}]$ denotes the \textit{adjacency matrix} between the $G$ nodes in the graph.
Our goal is to obtain an accurate estimate of $\textbf{A}$. To do that, 
we treat the indicators $\mathbf{b}_{mn}$ as missing data, and use the EM algorithm \citep{EM}
to estimate the parameters of the mixture model. 
The hierarchical and parsimonious nature of the $L_2N$ model leads to shrinkage estimation and borrowing power across all
pairs of genes, as well as to computational efficiency. This is critical, since $K$ is typically very large and can be much larger than the 
sample size, $N$. 
Details regarding the parameter estimation for the $L_2N$ model
can be found in \cite{BarSchifano}.
The algorithm is implemented as an R-package called \texttt{edgefinder} (see the Supplementary Material).

\subsection{Implementation Notes}\label{subsec.mixture.notes}
There are a couple of challenges related to the parameter estimation via the EM algorithm which should be addressed. First, the complete set of 
pairwise correlation coefficients, and thus the normalized weights, $w_{mn}$, cannot be assumed to
be mutually independent. Conditionally, they are all asymptotically normally distributed with variance $\frac{1}{N-3}$,
but for a fixed $m$, some $w_{mn}$ may
be correlated. Second, obtaining estimates based on all $K$ pairwise correlations 
is time-consuming and requires storing a very large matrix in the computer's memory, 
since the values of the indicator variables, $\hat{b}_{jmn}$, for each pair of genes must be 
updated in each iteration of the EM algorithm. 

When the number of genes is large, as is the case in the applications which motivated this paper,
we propose taking a random sample of $G'<G$ genes (e.g., $G'=1000$) and fitting the mixture model
to this random subset.
Using the smaller subset of genes greatly improves computational efficiency and yields highly accurate
estimates. Furthermore, randomly selecting a subset of the genes allows us to assume that the remaining
weights used in the (complete data) likelihood function are \textit{approximately} independent.
With the estimates obtained from the random subset, it is then possible to compute the posterior  
probabilities (\ref{eq:posterior_prob}) for all $K$ pairs, even on a computer with standard memory capacity.

In addition to computational efficiency and increasing power through shrinkage estimation, 
the mixture model
allows us to estimate how many pairs of genes are correctly (or incorrectly)
classified as co-expressed.
Specifically, 
for a predetermined posterior probability
ratio threshold, $T>1$, we solve
\begin{eqnarray*}
c_1 = \arg\min_{w\in (0,\infty)}\frac{\hat{p}_1\hat{f}_1(w)} {\hat{p}_0\hat{f}_0(w)} > T
\text{  \kern 1em and \kern 1em  }
c_2 = \arg\max_{w\in (-\infty,0)}\frac{\hat{p}_2\hat{f}_2(w)} {\hat{p}_0\hat{f}_0(w)} > T
\end{eqnarray*} 
and set $b_{0mn}=1$ if $w_{mn}\in [c_2, c_1]$, and $b_{0mn}=0$ otherwise.
Alternatively, for any $\alpha$, we can (numerically) find thresholds $c_1>0$ and $c_2<0$ such that 
\begin{eqnarray}
Pr(\hat{b}_{0mn}\ne 0~|~b_{0mn}= 0) &\approx&
   \hat{p}_0\int_{-\infty}^{c_2}\hat{f}_0(w)dw+\hat{p}_0\int_{c_1}^{\infty}\hat{f}_0(w)dw \le \alpha\,.\label{PPR1}
\end{eqnarray}
That is, we control the estimated probability of a Type I error at a certain level, $\alpha$. Similarly, we can control
the false discovery rate \citep{benjamini:hochberg}.
Using the thresholds $c_1$ and $c_2$, we can estimate the probability of a Type II error:
\begin{eqnarray}
Pr(\hat{b}_{0mn}= 0~|~b_{0mn}\ne 0) &\approx& \hat{p}_2\int_{c_2}^{0}\hat{f}_2(w)dw +\hat{p}_1\int_{0}^{c_1}\hat{f}_1(w)dw \label{PPR2}.
\end{eqnarray}

\section{Simulation Study}\label{simulations}
\subsection{Data Generated Under the $L_2N$ Model}\label{L2Nsim}
In the first simulation study, we assess the power and goodness of fit of our model using
different configurations, with varying numbers of genes (G), samples (N),
degrees of sparsity ($p_1+p_2$), and graph structures.
In this section, data are generated according to the $L_2N$ model from Section \ref{sec.model},
and we use different parameters for the log-normal components. 
We show representative results with $N=100$ and $G=500$ (thus, $K$, the maximum possible 
number of edges is 124,750.)
Four network configurations are used in this section. They are described in terms of the shape of the 
$G\times G$ adjacency matrix, $A$, as follows:
\begin{itemize}
\item \textbf{complete}: $A=BlockDiag(J_S-I_S, 0_{G-S})$, where $S=100$, $J$ is a matrix of 1's, $I$ is
an identity matrix, and $0$ is a matrix of zeros. 
This graph contains one clique (complete subgraph) with 100 nodes,
and nodes not in the clique are not connected to other nodes. $|E|= 4,950$ ($\approx 0.04K$), $p_1=0.0396$, $p_2=0$.
\item \textbf{ar} (autoregressive):  $A$ has a Toeplitz structure, with $A_{ij}=1/(1+|i-j|)$ if both $i,j \le S$, where $S=100$.
$|E|=  4,950$,  $p_1=0.0396$, $p_2=0$. 
\item \textbf{two independent blocks}: $A=BlockDiag(J_S-I_S, J_S-I_S, 0_{G-2S})$, where $S=50$. 
(I.e., two distinct cliques, each
with 50 genes.) $|E|=  2,450$,   $p_1=0.0196$, $p_2=0$.
\item \textbf{two negatively correlated blocks}: Similar to the previous configuration, but the
two blocks are negatively correlated. $|E|= 4,950$, $p_1=0.0196$, $p_2=0.02$.
\end{itemize}

For pairs $i,j$ such that $A_{ij}=0$, we generated $w_{ij}$ independently
from a standard normal distribution. In the \textit{complete} and \textit{two independent blocks}
configurations, for $A_{ij} = 1$ we generated only positively correlated pairs (so $p_2=0$), and in the 
\textit{two negatively correlated blocks} configuration, 
the pairs were positively correlated within each block, but pairs across the two blocks were generated to be negatively correlated.
For the \textit{ar} structure, weights generated from the log-normal distribution
appear in the off-diagonals of $A$ in decreasing order. 
That is, the largest $G-1$ weights
are placed randomly in the secondary diagonal (elements $A_{i,i+1}$), the next $G-2$ 
largest weights are placed randomly in the ternary diagonal (elements $A_{i,i+2}$), etc.
(Note that the
values $A_{ij}$ in this case are only used to indicate that elements along diagonals are equal, but
these values are not used to generate the weights.)

All four configurations are sparse, with only 2-4\% of the putative edges being present in the graph.
The \textit{two negatively correlated blocks} structure has the same sparsity as the \textit{complete}
and \textit{ar} graphs, but it has different weights of non-null mixture components. The \textit{ar}
graph has a more constrained structure, with weights among pairs of nodes which decay as 
$|i-j|$ increases, for $i,j \le S$. Recall, however, that our method does not rely on any assumptions about
the structure of the graph. Therefore, it is expected that its performance will only depend on the
parameters involved in the mixture model. In the simulations presented here we examined the power
of the method to detect edges in the graphs as a function of the location parameters of the log-normal
components.  We varied $\theta_1$, such that $\theta_1\in\{-1.25, -1, -0.75,\ldots,0.75\}$, and 
set $\kappa_1^2=0.25$. In the
\textit{two negatively correlated blocks} configuration, we used $\theta_2=\theta_1$ and $\kappa_2=\kappa_1$.
The weights which were  generated according to the $L_2N$ model were transformed into correlation coefficients using the $\tanh$ function,
and the resulting covariance matrix was used to simulate 500 gene expression values for 100
subjects. 
For each configuration we generated 20 different datasets.

\begin{figure}
	\centering
	\includegraphics[width=0.5\textwidth]{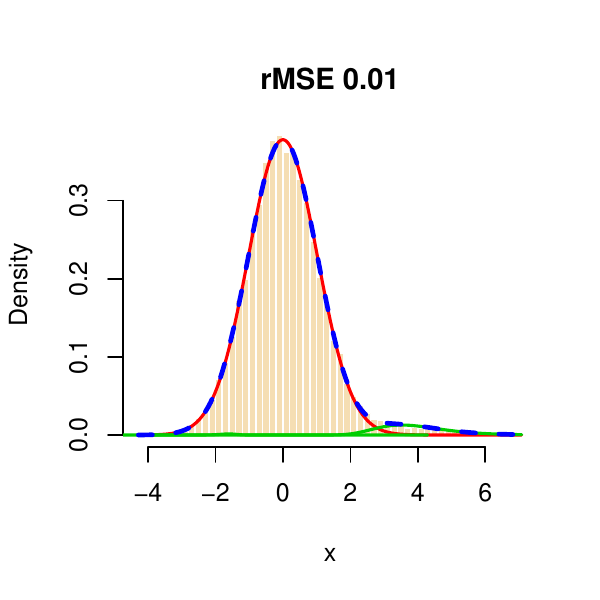}
	\caption{The distribution of $w_{mn}=arctanh(r_{mn})$ for a simulated dataset with 500 genes,
	of which 100 form a complete subgraph. The total number of edges in the graph is 4,950 (out of
	a total of 124,750 possible edges.)
	The red curve represents the null component, the green curves represent the non-null components,
	and the dashed blue line represents the fitted mixture distribution.}\label{fig.sim.clique}
\end{figure}

We applied our method and checked the goodness of fit 
of the mixture model and the ability to correctly recover the structure of the network, in terms
of the number of true- and false-positive edges. In our simulations, we used the approach
described in Section \ref{sec.model} to control the false discovery rate at the 0.01 level.
To demonstrate how well our algorithm estimates the true mixture model, we plotted
for each configuration the histogram of the observed $w_{mn}$ and the fitted 
mixture 
and measured the goodness of fit in terms of the root means squared error
(rMSE). In all configurations, the rMSE was very small
($\le 0.01$), and the mixture weights ($p_0$, $p_1$, $p_2$) were estimated
very accurately and with increasing accuracy as $\theta_1$ increases.
See, for example, Figure \ref{fig.fittedp1} in the supplementary material, where the average
estimate of $p_1$ is plotted versus $\theta$ using the \textit{two negatively correlated blocks}
configuration.
A representative goodness of fit plot is shown in Figure \ref{fig.sim.clique},
for the \textit{complete} configuration, with $\theta_1=-0.25$. The red curve represents the null
component, the green lines represent the non-null components (in this case, $C_2$ is estimated,
correctly, to have a weight which is very close to 0), and the dashed blue line is the mixture.

\begin{figure}
	\centering
	\includegraphics[trim={0 0.6cm 0 1cm},clip,scale=0.7]{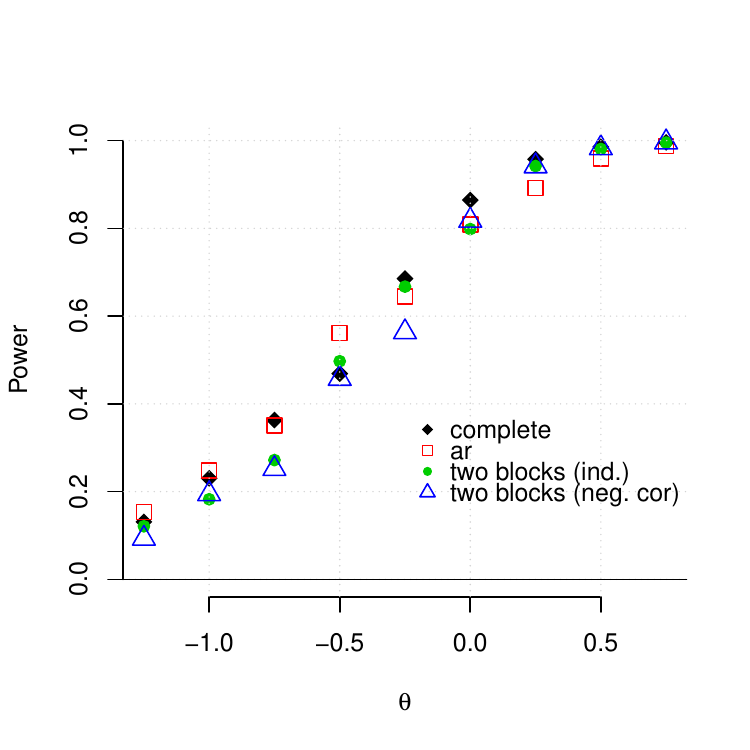}
	\caption{The average power of our method when the data are generated according to the $L_2N$
		mixture model, as a function of $\theta$, for a simulated dataset with 500 genes
		and four different forms of adjacency matrices (complete, autoregressive, two independent blocks, and
		two negatively correlated blocks).
		In all cases, the FDR was controlled at the 0.01 level.}\label{fig.sim.clique.power}
\end{figure}

Arguably, more important than assessing goodness of fit, is determining a method's 
ability to recover the true network structure correctly, i.e. identify as many existing
edges as possible while maintaining a low number of falsely-detected edges. Figure \ref{fig.sim.clique.power} shows the average power of our method 
to detect true edges for
a range of values for $\theta_i$ (with a fixed $\kappa^2$) and for different network
configurations, where power is the
total number of true-positives divided by the total number of edges in the graph.
It can be seen that, as the location parameter of the log-normal distribution increases, the 
power increases and approaches 1.
When the data are generated under the $L_2N$ model, this is the expected behavior,
 since as $\theta_1$ ($\theta_2$)
increases, the positive (negative) non-null component, $C_1$ ($C_2$), is pushed further to the right (left),
making it easier to discriminate between non-null and null components.
Note that our method has approximately the same power curve for all four configurations.

\begin{figure}[h]
	\centering
	\includegraphics[trim={0 0.6cm 0 1.2cm},clip,scale=0.8]{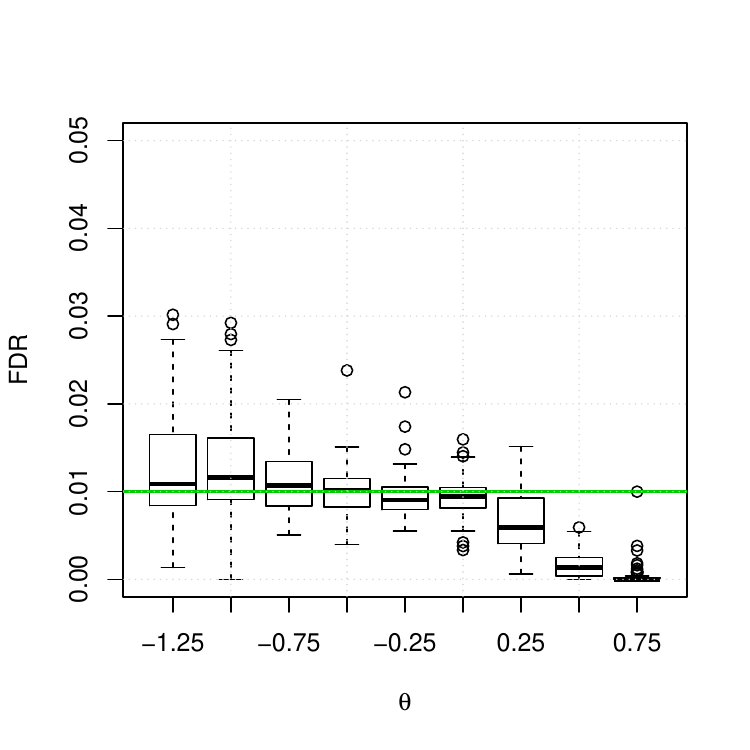}
	\caption{The observed false discovery rate in the simulations under the $L_2N$ model (Section \ref{L2Nsim}), as a function of the location parameter of the log-normal distributions of the non-null components.
	The green horizontal line shows the level (0.01) which we used to control the false discovery rate. Using
	the notation Section \ref{sec.model}, we determined the thresholds, $c_1$ and $c_2$, that correspond to this
	FDR level. The boxplots show the distributions across four network structures, with 20 replications in each configuration.}\label{fig.fittedFDR}
\end{figure}

The average false discovery rate across all configurations and replications is 0.008.
For small values of $\theta_1$ ($< -0.75$), the average FDR is slightly higher (approximately 0.012) and for
$\theta>0$, the average FDR is less than 0.01. More detailed results regarding the
achieved false discovery rate are shown graphically in Figure
\ref{fig.fittedFDR}.

\subsection{Data Generated Under Other Models}
In the second simulation study we evaluate the ability of our method to recover the true network structure and compare it with other
methods. For a fair comparison, the data are not generated under $L_2N$ model.
Rather, we use data generated from a multivariate normal distribution whose covariance matrix $\Sigma$ is defined as a function of an adjacency matrix $A$. To generate data, we use the \textsf{R}-package \textsf{huge} \citep{rpackagehuge}. 
We generate five types of network configurations as follows: 
\begin{itemize}
\item  \textbf{random}: each edge is randomly set to exist in the graph using $K$ i.i.d.
Bernoulli(p) draws ($p = 0.01, 0.05, 0.1$). $|E|\approx 1000 \times (1000 - 1) \times p / 2$.
\item  \textbf{hub}: it consists of $g$ disjoint groups, and nodes within each group are
only connected through a central node in the group ($g = 25, 50, 100$). $|E|=1000 - g$.
\item \textbf{band}: an edge, $e_{mn}$ between nodes $m\ne n$ is set to exist in the graph if 
$1 \leq |m-n| \leq g$ ($g = 25, 50, 100$). $|E|=(2000 - 1 - g)\times g / 2$.
\item  \textbf{scale-free}: scale-free networks are generated using the Barab\'{a}si-Albert algorithm \citep{barabasi1999emergence}. $|E|=1,000$.
\item \textbf{overlapped-cluster}: 
we modified a function that generates a \textit{cluster} network that consists of non-overlapping $g$ groups. In the modified function, the groups are aligned in an adjacency matrix so that each group shares 20\% of the nodes with its left-adjacent group and another 20\% with its right-adjacent group. Edges in each group are randomly generated with probability $p$. We use $p = 0.3$ for $g = 25$ and $50$, and $p = 0.6$ for $g = 100$. $|E| \approx (0.8 \times (g - 1) + 1) \times (1000/g) \times (1000/g-1) \times p / 2 $. Edges in each group are randomly generated with probability $p$.
\end{itemize}
For each configuration, we generated expression profiles of $G=1,000$ genes for $N=70$ samples. We compare our method with three existing methods: Meinshausen-B{\"u}hlmann graph estimation (\textbf{MB}, \cite{meinshausen2006high}), graphical lasso (\textbf{glasso}, \cite{friedman2008sparse}), and \textbf{correlation thresholding} graph estimation. 
For each of MB and glasso, we identify edges in two different ways. First, an edge $e_{mn}$ between nodes $m$ and $n$ is estimated to exist (i.e. $\hat{A}_{mn} = 1$) if the method chooses the node $m$ as a neighbor of $n$ \textbf{and} the node $n$ as a neighbor of $m$, which we name as MB-AND and glasso-AND, respectively. Second, an edge $e_{mn}$ is estimated to exist if the method chooses the node $m$ as a neighbor of $n$ \textbf{or} the node $n$ as a neighbor of $m$, which we name as MB-OR and glasso-OR, respectively. The sparsity levels of MB and glasso are controlled by a regularization parameter $\lambda$ and the correlation thresholding method by setting different thresholds from 0 to  three times the true sparsity level.
%

We compare the performance in terms of the number of true positives (i.e. correctly identified edges) \textit{given the \textbf{same} total number of edges identified}. We define the true network in two different ways: (i) based on the adjacency matrix --- a pair $m,n$ is set to be connected if and only if $A_{mn}=1$, and (ii) by applying a threshold to the true covariance matrix --- for a predetermined $t$, a pair is connected iff $|\Sigma_{mn}|>t$ where the threshold $t$ is set to achieve the same sparsity level as that of the adjacency matrix. The two are different in that the former assumes conditional independence between two nodes that not connected by an edge, while the latter does not. 

The  results from the two network definitions are similar. We show the result from (i) whose true matrix is derived from the adjacency matrix, and show the result from (ii) whose true matrix is derived from the true covariance matrix in the supplementary material (Figure~\ref{fig.sim.TPcovmat}).

\begin{figure}
	\centering
	\includegraphics[width=0.9\textwidth]{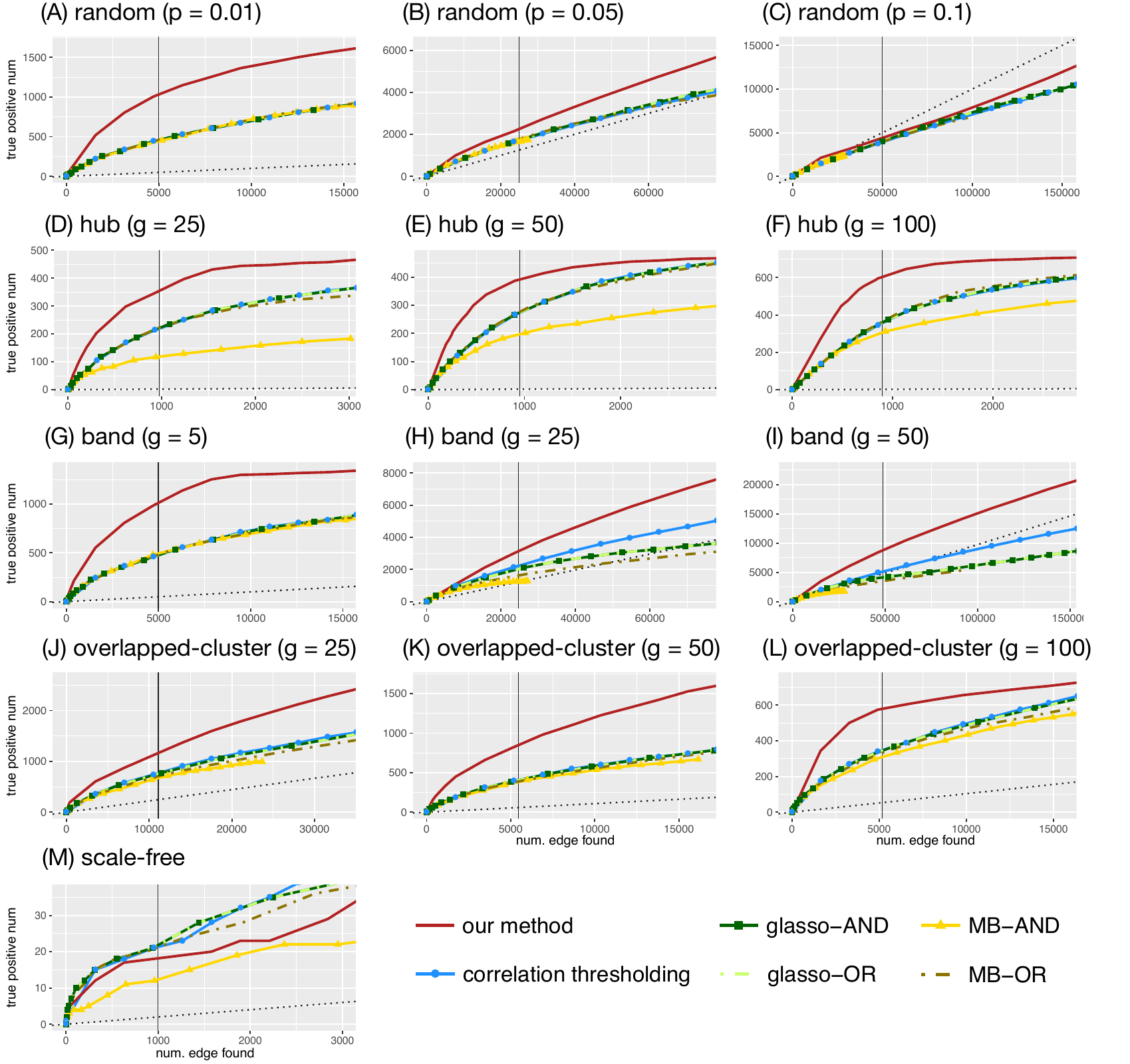}
	\caption{The numbers of true positive edges given the total number of edges identified by each method. The adjacency matrix used by the \textsf{huge} package is used to determine the true edges. The y-axis represents the number of true positive edges and the x-axis represents the total number of edges identified. The vertical line represents the number of true edges. The black dotted line is a regression line with 0 intercept and slope equal to the true sparsity, which represents the expected number of true positive edges when the edges are identified in a random manner (uniformly).
	}\label{fig.sim.TPadjmat}
\end{figure}

Figure~\ref{fig.sim.TPadjmat} depicts the numbers of true positive edges given the total 
number of edges identified by each method. Our method (solid red lines) outperforms all the other 
methods in random, hub, band, and overlapped-cluster networks (see Figure~\ref{fig.sim.TPadjmat}A-L). For example, in the hub, $g=100$ configuration (panel F), the true number of edges is 900 (as indicated by the vertical line), and when our method detects 853 edges, 595 of them are true edges. In contrast, the thresholding method yields approximately 345 true edges out of the total 856 detected; MB-OR and glasso- yield approximately 370--380 true edges out of 900--960 edges detected; and MB-AND yields 312 true edges out of 930 edges detected. Notably, MB and glasso are comparable to, but in some cases worse than the correlation thresholding method in all network configurations. 

The black dotted line in each plot represents the expected number of true positive edges
when the edges are identified in a random manner. In the band, $g=50$ configuration (panel I), the competing methods do as well as or worse than chance, whereas our method gives much better results.
In the random, $p=0.1$ configuration (panel C), the other methods perform worse than choosing
edges randomly and our method performs slightly better than chance to a certain point (approximately 30,000 total edges being detected) before deteriorating, but still gives better results than the other methods.

The scale-free configuration appears to be especially challenging for all the methods. They all detect only approximately 10 -- 20 true edges when the total number of edges detected was 1,000 (Figure~\ref{fig.sim.TPadjmat}M). In this case, the correlation thresholding, glasso- and MB-OR perform only slightly better than our method.

\begin{figure}[h]
	\centering
	\includegraphics[width=0.9\textwidth]{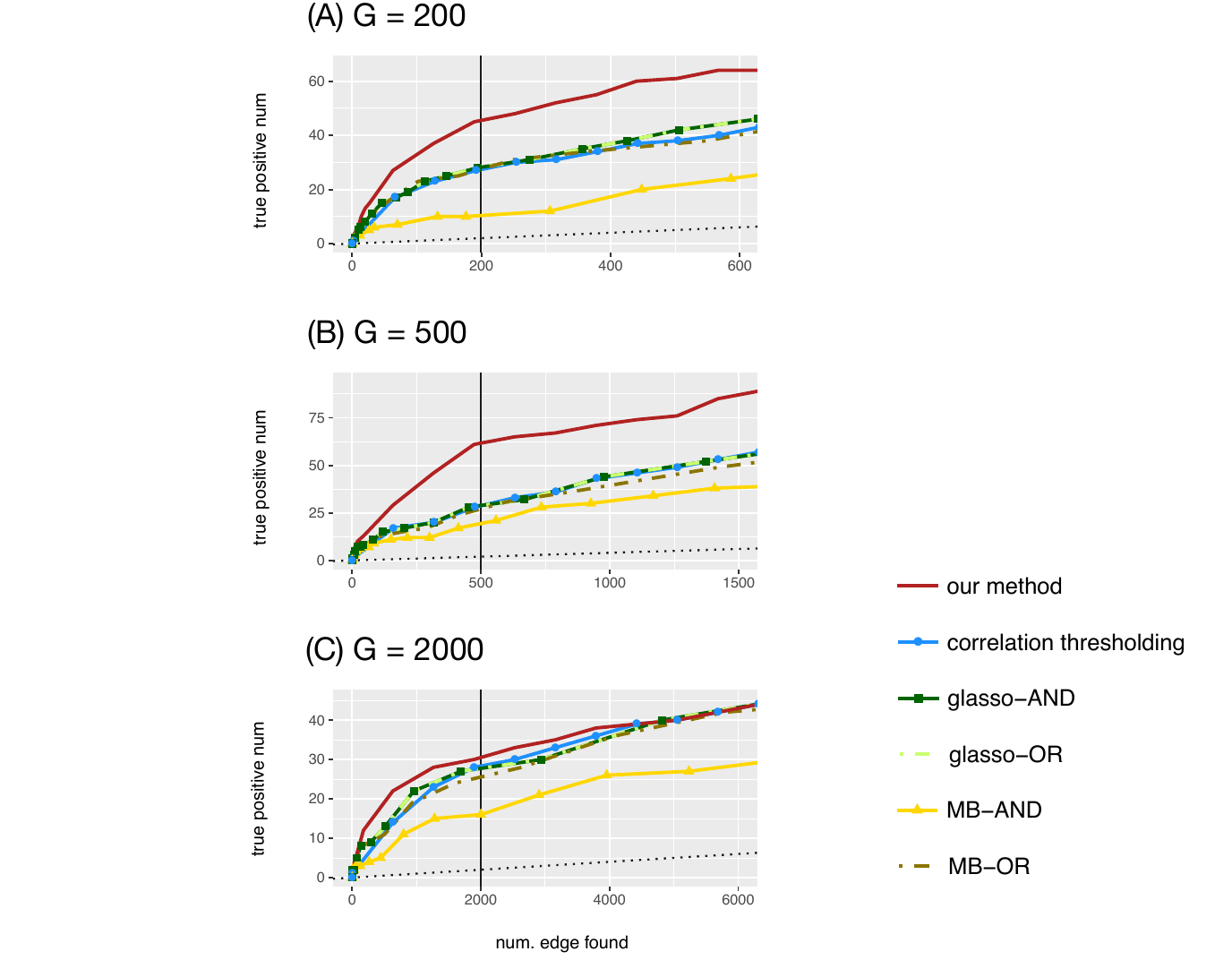}
	\caption{The numbers of true positive edges given the total number of edges using three scale-free network configurations ($G = 200, 500, 2000$, where $G$ is the number of genes). The adjacency matrix is used as the true matrix. The y-axis represents the number of true positive edges and the x-axis represents the total number of edges identified. The vertical line represents the number of true edges. The black dotted line is a regression line with 0 intercept and slope equal to the true sparsity, which represents the expected number of true positive edges when the edges are identified in a random manner.}\label{fig.sim.TPadjmat.scalefree}
\end{figure}

We further explore scale-free networks with additional network sizes  ($G = 200, 500, 2000$). Consider that sparsity levels change as we vary the network sizes: larger $G$ results in a more sparse network. Figure~\ref{fig.sim.TPadjmat.scalefree} depicts the numbers of true positive edges given the total number of edges for various sizes of scale-free networks. When $G = 2,000$, our method is much better than MB-AND (as we have also seen in Figure~\ref{fig.sim.TPadjmat}M), and slightly better than the other methods. However, when $G = 200, 500$, our model outperforms the other methods. For example, when $G = 200$, our model identifies around 45 true edges out of 200 edges, while glasso-, MB-OR, and the correlation thresholding methods identify less than 30 true edges, and MB-AND identifies around 10 edges.

\section{Case Study}\label{sec:casestudy}
\textbf{Subtype Characterization of Breast Cancer.}
We analyzed a large dataset used by \citet{Allahyar2019} who introduced SyNet, a computational tool aiming 
to improve network-based cancer outcome prediction. The dataset, referred to as ACES \cite{ACES}, contained expression
data for 12,750 genes collected from 1,616 women across 12 studies. 
To obtain and preprocess the data, we used the raw data and scripts provided by \citet{Allahyar2019} 
via their github page (\url{github.com/UMCUGenetics/SyNet}).
We focused on analyzing gene networks by cancer subtype: Basal (n=297), Her2 Positive (n=191),
Luminal A (n=584), Luminal B (n=440), and Normal (n=104).
Categorization into  breast cancer types is based on the positive/negative status of three factors: estrogen receptor (ER),
progesterone receptor (PR), and the number of copies of the HER2 gene.
The Basal group is called the triple-negative breast cancer, and it includes tumors in which
all three markers (ER, PR, HER2) are negative.
The Her2 group includes tumors that are ER and PR negative, but HER2 positive. The Luminal A group includes tumors that are ER positive and PR positive, but HER2 negative. The Luminal B group includes tumors that are ER positive, PR negative, and HER2 positive.

We selected 80 random samples from each group, and applied \texttt{edgefinder} to detect the edges in
the network in each of the five groups. We controlled the false discovery rate at the 0.01 level.
With 12,750 genes there are 81,274,875 possible edges in the network, and  \texttt{edgefinder} detects
112,071 edges in the network of the Normal group, 171,959 in the Basal group,
144,128 in the Her2 group, 278,066 in Luminal A, and 213,200 in Luminal B. 
Clearly, all five networks are very sparse  (0.1-0.3\% sparsity). All but two edges (in the Basal network)
belonged to the positive correlation component in the mixture model. 
The results are very stable, in the sense of \citet{meinshausen2006high},
so that when we picked 100 different
subsamples from each subtype, the same edges were detected in each of the cancer groups, and
in the normal group the same network was detected 99 times. The mixture model for the normalized 
correlation coefficients fits all five very well, with rMSE=0.01. For example, Figure S5
in the Supplementary Material shows the fitted curve for the Basal group.

\begin{figure}[ht]
	\hfill
	\subfigure[Normal]{\includegraphics[width=8cm]{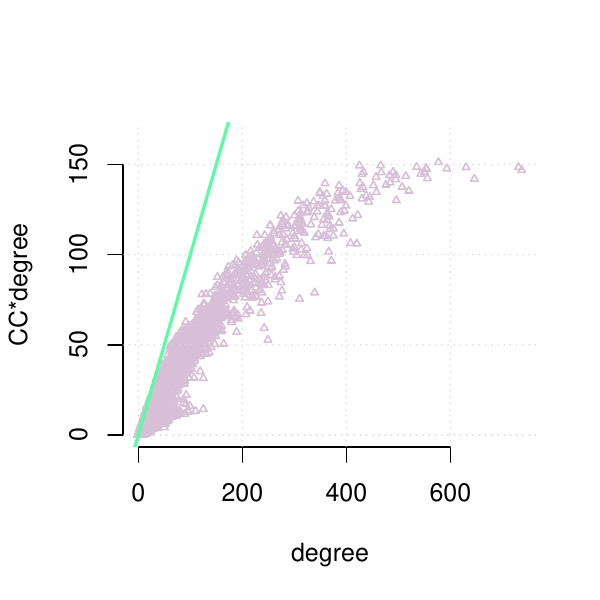}}
	\hfill
	\subfigure[Her2]{\includegraphics[width=8cm]{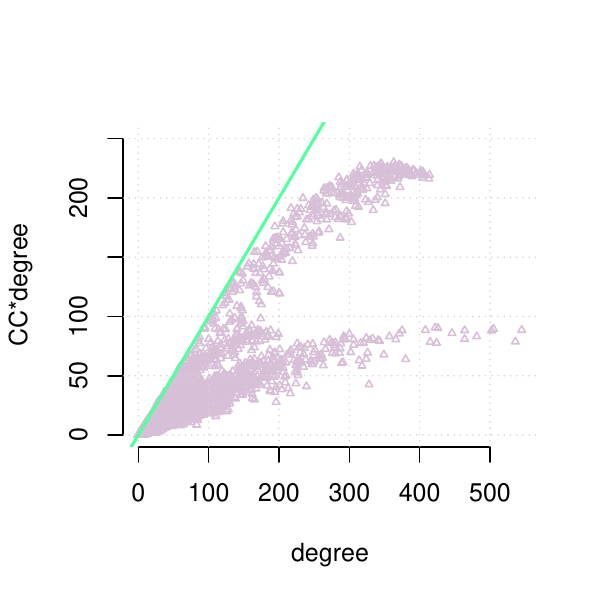}}
	\hfill
	\caption{Plot of $\gamma_m d_m$ versus $d_m$, where $\gamma_m$ is the clustering coefficient and  $d_m$ is the degree of node $m$.}\label{fig.degcc}
\end{figure}

\begin{figure}[ht]
	\centering
	\includegraphics[width=0.5\textwidth]{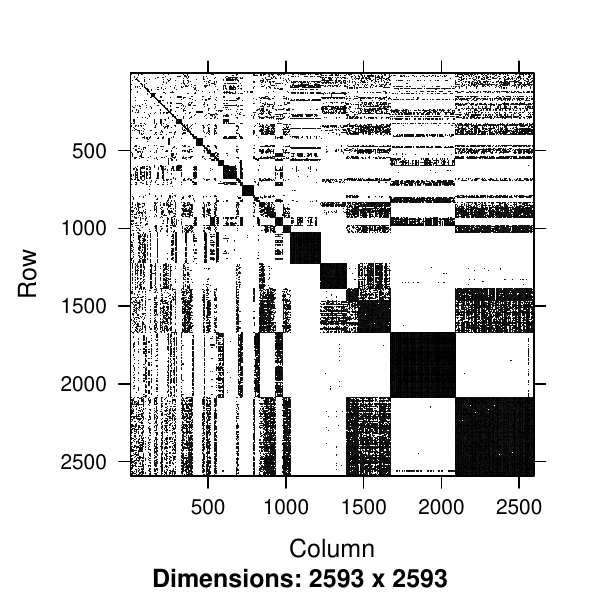}
	\caption{A bitmap of the Her2 network, including only genes with at least 20 neighbors.
		A black dot represents an edge between a pair of nodes, and white dots represent pairs which are not connected. The genes were ordered according to their degree, within each cluster.}\label{fig.CCimage}
\end{figure}

We now consider two  characteristics of nodes, namely the degree, $d_m$, and the clustering coefficient, $\gamma_m$, where $m$ indicates a node $m$. Let $N_m=\{n~|~A_{mn}=1\}$ be the set of nodes adjacent to node $m$, and let $E(N_m)$ be the set of edges between nodes in $N_m$. Then
\begin{eqnarray*}
	d_m = |N_m|\text{  \kern 1em and \kern 1em  }
	\gamma_m = \frac{|E(N_m)|}{d_m(d_m-1)/2}\,.
\end{eqnarray*}
If $d_m\le 1$, the clustering coefficient is defined as $\gamma_m=0$. By definition, $\gamma_m\in[0,1]$. The degree of a node is interpreted as the involvement of the node in the network and the clustering coefficient as the connectivity among neighbors of the node. Note that $\gamma_m d_m$ is, by definition, proportional to $|E(N_m)|/(d_m-1)$, so it is interpreted as (approximately) the average degree among the neighbors of node $m$. Note
also that $\gamma_m d_m$ is bounded by $d_m$. 
To visualize and compare the properties of
the networks, we plot $\gamma_m d_m$ versus $d_m$. 
For example,  Figure \ref{fig.degcc} depicts the relationship between the degree and the clustering
coefficients of nodes for (a) the Normal group and (b) the Her2 group. Similar plots for the other three subtypes are provided in the Supplementary Material.
In all five groups we observe that as the degree of a node increases, the average 
degree of each of its neighbors increases. 
The plot for the Her2 group has a striking pattern, with three distinct `arms', one of which 
exhibits a very high degree of connectivity between nodes (as shown by the high values of
$\gamma_m d_m$.) This suggests that there is at one cluster which is nearly a complete subgraph
in which each node is connected to all or most other nodes in that cluster.
The presence of such clusters is also depicted in Figure \ref{fig.CCimage}, which shows
a bitmap of the data (restricted to genes with at least 20 neighbors), where a black dot represents an 
edge between a pair of nodes and white dots
represent pairs which are not connected.
To detect clusters and to generate plots such as Figure \ref{fig.CCimage}, where clusters are
arranged by their size, we first define the dissimilarity between two neighboring nodes, $m$ and $n$,
as the number of neighbors of $m$ which are not neighbors of $n$, plus the number of 
neighbors of $n$ which are not neighbors of $m$. Then, we create clusters recursively as follows.
In each iteration, $i$, we find the node, $x_i$, with the largest degree and identify all of its neighbors.
We sort its neighbors in increasing dissimilarity order. So, within the $i$-th cluster, the
nodes are arranged from closest to $x_i$ to the farthest. Then, in the subsequent iterations node $x_i$ and
all its neighbors are excluded, until we have exhausted all the nodes.

\begin{figure}[ht]
	\centering
	\includegraphics[width=0.5\textwidth]{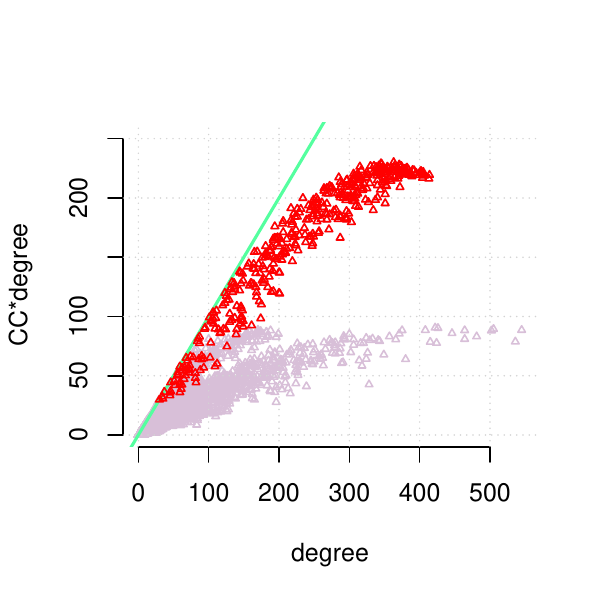}
	\caption{Plot of $\gamma_m d_m$ versus $d_m$ for the Her2 group. Points in red correspond to genes
		identified as belonging to the second largest cluster.}\label{fig.Her2Cluster2}
\end{figure}

Figure \ref{fig.CCimage} shows that the genes in the second largest cluster in the 
Her2 network appear to be highly connected. Figure \ref{fig.Her2Cluster2} shows again 
the plot of $\gamma_m d_m$ versus $d_m$, but this time, genes identified as belonging to the
second largest cluster appear in red. Clearly, the upper `arm' in the plot consists of genes from
that cluster.

\begin{table}[ht]
	\centering
	\caption[caption]{Pathways characterizing the breast cancer subtypes.} \label{table:pathwaylistcancer}
	\begin{adjustbox}{max width=\textwidth}
		\def\arraystretch{0.95}
		\begin{threeparttable}
			\begin{tabular}{@{}cccccp{13cm}@{}}
				\toprule
				Subtype & Pathway Name & Pathway Size & Cluster Index & P-value  & Genes \cr\midrule
				Her2    & Steroid hormone biosynthesis           & 42           & 18            & 0.000780 & AKR1C1, AKR1C2, HSD3B1, SRD5A1, AKR1D1, UGT2B28 \cr
				& Tryptophan metabolism                  & 35           & 18            & 0.002258 & KYAT1, DDC, MAOA, KYNU, HAAO \cr
				& Drug metabolism - cytochrome P450      & 54           & 18            & 0.009187 & ADH1C, CYP2A6, FMO5, MAOA, UGT2B28 \cr\midrule
				Basal   
				& RNA transport                          & 126          & 46            & 0.004090 & CLNS1A, RAN, RANBP2, NUP160, SUMO4 \cr
				& mRNA surveillance pathway              & 68           & 70            & 0.004276 & PPP2CA, PPP2R5C, PPP2R5E, PPP2R3C \cr\midrule
				LumA    & Bacterial invasion of epithelial cells & 65           & 3             & 0.006240 & CAV1, CAV2, HCLS1, ILK, ITGA5, PIK3CD, ELMO1, ARPC1B \cr\midrule
				Lum B   
				& Tight junction                         & 118          & 39            & 8.34E-05 & ACTN2, ACTN3, MYH1, MYH2, MYL2, MYLPF \cr
				& Base excision repair                   & 31           & 37            & 0.001882 & FEN1, PCNA, UNG \cr
				& Dilated cardiomyopathy                 & 83           & 42            & 0.002040 & MYL2, TNNC1, TPM3, TTN  \cr
				& Hypertrophic cardiomyopathy (HCM)      & 77           & 42            & 0.002040 & MYL2, TNNC1, TPM3, TTN \cr
				& Regulation of autophagy                & 30           & 3             & 0.003573 & IFNA4, IFNA10, IFNA17, IFNA21  \cr
				& Nucleotide excision repair             & 41           & 37            & 0.003760  & PCNA, RFC4, RFC5 \cr
				\bottomrule                                 
			\end{tabular}
			\begin{tablenotes}
				\footnotesize
				\item Uniquely enriched pathways in each breast cancer subtype. All identified clusters are indexed by their size. The cluster index indicates the order of the clusters. P-values are adjusted to control the false discovery rate using \citet{benjamini:hochberg}.
			\end{tablenotes}
		\end{threeparttable}
	\end{adjustbox}	
\end{table}

Next, we characterize each subtype by pathways enriched in its gene clusters. For each subtype, we perform gene set enrichment analysis \cite{subramanian2005gene} and find the KEGG pathways significantly enriched in every identified cluster (p-value $\leq 0.01$). We then identify uniquely enriched pathways for each subtype compared to other subtypes (see Table~\ref{table:pathwaylistcancer}). 

Our results obtained from the pathway analysis are consistent with the previous findings. The HER2 group has three uniquely enriched pathways: \textit{drug metabolism--cytochrome P450}, \textit{steroid hormone biosynthesis}, and \textit{tryptophan metabolism}. The relationship of \textit{drug metabolism--cytochrome P450} to HER2 has been reported in \citet{towles2016cytochrome}, the relationship of \textit{steroid hormone biosynthesis} to HER2 has been reported in \citet{huszno2015influence}, and the relationship of \textit{tryptophan metabolism} to HER2 has been reported in \citet{fisher2010structure} and \citet{miolo2016pharmacometabolomics}. The basal group has two uniquely enriched pathway that are related to genetic information processing for RNA translation: \textit{mRNA surveillance pathway} and \textit{RNA transport}. The luminal A group has one uniquely enriched pathway: \textit{Bacterial invasion of epithelial cells}. The luminal B group has six uniquely enriched pathways: \textit{Base excision repair}, \textit{Dilated cardiomyopathy}, \textit{Hypertrophic cardiomyopathy (HCM)}, \textit{Nucleotide excision repair}, \textit{Regulation of autophagy}, and \textit{Tight junction}. Among those, \textit{regulation of autophagy} \cite{cook2011autophagy}, \textit{base excision repair}, and \textit{nucleotide excision repair} \cite{anurag2018comprehensive} are associated with endocrine treatment resistance in breast cancer patients characterized by the expression of estrogen receptor alpha. The normal group is characterized by pathways related to the immune system while those pathways are not enriched or enriched in fewer clusters in all breast cancer subtypes. \textit{Endocytosis} and \textit{lysosome} that are uniquely enriched in the normal group are parts of the  immune system. Furthermore, \textit{Phagosome}, \textit{antigen processing and presentation}, \textit{natural killer cell mediated cytotoxicity}, \textit{graft-versus-host disease}, \textit{T cell receptor signaling pathway}, and \textit{primary immunodeficiency} are enriched in a larger number of clusters in the normal group than other subtypes. These pathways are also part of or related to the immune system (see Table~\ref{table:pathwaylistnormal} in the Supplementary Material).

\section{Discussion}\label{discussion}
We propose a new approach for detecting edges in gene networks, based on
co-expression data. We consider the entire set of genes as a network in which nodes represent 
genes and weights on edges represent the correlation between expression
levels of pairs of genes. We start by modeling the normalized pairwise correlations as a mixture of three components: a normal component with mean 0, representing the majority of pairs which are not co-expressed, and two non-null components, modeled as log-normal distributions, for positively and negatively correlated pairs.

From a theoretical point of view, the so-called $L_2N$ model has the advantage that the
overlap between the null component and the non-null components around 0 is negligible.
This helps to avoid identifiability problems which are known to affect other mixture models which
rely only on normal components, such as the spike-and-slab or a three-way normal mixture model.
Furthermore, the mixture model allows us to accurately
estimate the proportion of spurious correlations among all pairs of genes and to 
derive a cutoff criterion in order to eliminate the vast majority of the edges
in the graph that correspond to uncorrelated genes. We also derived
estimators for the probabilities of Type-I and Type-II errors, as well as the false discovery rate
associated with the 
cutoff criterion.

From a practical point of view, this model appears to fit co-expression data extremely well,
 even when the data are not generated according to the mixture model. Our
 simulations show that it outperforms other methods in terms of the power to detect edges, and that
 it maintains a low false discovery rate. Estimation of the model parameters is done very
 efficiently, using the EM algorithm. In typical gene expression datasets 
 which consist of thousands of genes and millions of putative edges, 
 computational
 efficiency is critical.

Our approach does not require any assumptions about the underlying structure of the network. We only assume that the normalized correlations follow the $L_2N$ model. This is a very
modest assumption since  the Fisher z-transformed correlations are indeed (asymptotically) normally distributed for all the uncorrelated pairs.

%
Our case study yielded results that are insightful and consistent with previous findings. In particular, for the HER2 group, we identified the pathways that are known to be associated with this subtype. For luminal B group, we identified pathways that are associated with endocrine treatment resistance in breast cancer with estrogen receptor alpha. All the breast cancer subtypes are less prone to have pathways related to the immune system than the normal group, which is reasonable because cancer is known to weaken the immune system. 

We plan to extend this method to handle time varying networks. This will be particularly useful
when analyzing gene expression data from repeated measures designs. We also plan to extend the model
to applications that involve multiple platforms, such as methylation and proteomics. In
principle, with the appropriate normalization technique for each platform, one can simply
construct a graph with $G_1+\ldots +G_k$ nodes, where $G_i$ is the number of `building blocks' observed in each platform. However, much more work is needed to establish the theoretical framework to define `co-expression' across platforms.

%


\subsection*{Supplementary Material}
%
%
See attached document for additional figures for the simulations and case study. 
The \texttt{edgefinder} R-package is available from the first author's website \url{ https://haim-bar.uconn.edu/software/}.



\bibliographystyle{unsrtnat}
\bibliography{references}

\begin{thebibliography}{47}
\providecommand{\natexlab}[1]{#1}
\providecommand{\url}[1]{\texttt{#1}}
\expandafter\ifx\csname urlstyle\endcsname\relax
  \providecommand{\doi}[1]{doi: #1}\else
  \providecommand{\doi}{doi: \begingroup \urlstyle{rm}\Url}\fi

\bibitem[Stuart et~al.(2003)Stuart, Segal, Koller, and Kim]{stuart2003gene}
Joshua~M Stuart, Eran Segal, Daphne Koller, and Stuart~K Kim.
\newblock A gene-coexpression network for global discovery of conserved genetic
  modules.
\newblock \emph{science}, 302\penalty0 (5643):\penalty0 249--255, 2003.

\bibitem[Zhang and Horvath(2005)]{zhang2005general}
Bin Zhang and Steve Horvath.
\newblock A general framework for weighted gene co-expression network analysis.
\newblock \emph{Statistical applications in genetics and molecular biology},
  4\penalty0 (1), 2005.

\bibitem[Eisen et~al.(1998)Eisen, Spellman, Brown, and
  Botstein]{eisen1998cluster}
Michael~B Eisen, Paul~T Spellman, Patrick~O Brown, and David Botstein.
\newblock Cluster analysis and display of genome-wide expression patterns.
\newblock \emph{Proceedings of the National Academy of Sciences}, 95\penalty0
  (25):\penalty0 14863--14868, 1998.

\bibitem[Taylor et~al.(2009)Taylor, Linding, Warde-Farley, Liu, Pesquita,
  Faria, Bull, Pawson, Morris, and Wrana]{taylor2009dynamic}
Ian~W Taylor, Rune Linding, David Warde-Farley, Yongmei Liu, Catia Pesquita,
  Daniel Faria, Shelley Bull, Tony Pawson, Quaid Morris, and Jeffrey~L Wrana.
\newblock Dynamic modularity in protein interaction networks predicts breast
  cancer outcome.
\newblock \emph{Nature biotechnology}, 27\penalty0 (2):\penalty0 199--204,
  2009.

\bibitem[Barab{\'a}si et~al.(2011)Barab{\'a}si, Gulbahce, and
  Loscalzo]{barabasi2011network}
Albert-L{\'a}szl{\'o} Barab{\'a}si, Natali Gulbahce, and Joseph Loscalzo.
\newblock Network medicine: a network-based approach to human disease.
\newblock \emph{Nature Reviews Genetics}, 12\penalty0 (1):\penalty0 56--68,
  2011.

\bibitem[Ravasz et~al.(2002)Ravasz, Somera, Mongru, Oltvai, and
  Barab{\'a}si]{ravasz2002hierarchical}
Erzs{\'e}bet Ravasz, Anna~Lisa Somera, Dale~A Mongru, Zolt{\'a}n~N Oltvai, and
  A-L Barab{\'a}si.
\newblock Hierarchical organization of modularity in metabolic networks.
\newblock \emph{science}, 297\penalty0 (5586):\penalty0 1551--1555, 2002.

\bibitem[Wuchty et~al.(2003)Wuchty, Oltvai, and
  Barab{\'a}si]{wuchty2003evolutionary}
Stephan Wuchty, Zolt{\'a}n~N Oltvai, and Albert-L{\'a}szl{\'o} Barab{\'a}si.
\newblock Evolutionary conservation of motif constituents in the yeast protein
  interaction network.
\newblock \emph{Nature genetics}, 35\penalty0 (2):\penalty0 176--179, 2003.

\bibitem[Tong et~al.(2004)Tong, Lesage, Bader, Ding, Xu, Xin, Young, Berriz,
  Brost, Chang, et~al.]{tong2004global}
Amy Hin~Yan Tong, Guillaume Lesage, Gary~D Bader, Huiming Ding, Hong Xu,
  Xiaofeng Xin, James Young, Gabriel~F Berriz, Renee~L Brost, Michael Chang,
  et~al.
\newblock Global mapping of the yeast genetic interaction network.
\newblock \emph{science}, 303\penalty0 (5659):\penalty0 808--813, 2004.

\bibitem[Newman(2001)]{newman2001scientific}
Mark~EJ Newman.
\newblock Scientific collaboration networks. ii. shortest paths, weighted
  networks, and centrality.
\newblock \emph{Physical review E}, 64\penalty0 (1):\penalty0 016132, 2001.

\bibitem[Amaral et~al.(2000)Amaral, Scala, Barthelemy, and
  Stanley]{amaral2000classes}
Lu{\i}s A~Nunes Amaral, Antonio Scala, Marc Barthelemy, and H~Eugene Stanley.
\newblock Classes of small-world networks.
\newblock \emph{Proceedings of the national academy of sciences}, 97\penalty0
  (21):\penalty0 11149--11152, 2000.

\bibitem[Jeong et~al.(2001)Jeong, Mason, Barab{\'a}si, and
  Oltvai]{jeong2001lethality}
Hawoong Jeong, Sean~P Mason, A-L Barab{\'a}si, and Zoltan~N Oltvai.
\newblock Lethality and centrality in protein networks.
\newblock \emph{Nature}, 411\penalty0 (6833):\penalty0 41--42, 2001.

\bibitem[Zhang et~al.(2011)Zhang, Watson, and Heath]{zhang2011network}
Liqing Zhang, Layne~T Watson, and Lenwood~S Heath.
\newblock A network of scop hidden markov models and its analysis.
\newblock \emph{BMC bioinformatics}, 12\penalty0 (1):\penalty0 191, 2011.

\bibitem[Chu et~al.(2012)Chu, Rivera, Popel, and Bader]{chu2012constructing}
Liang-Hui Chu, Corban~G Rivera, Aleksander~S Popel, and Joel~S Bader.
\newblock Constructing the angiome: a global angiogenesis protein interaction
  network.
\newblock \emph{Physiological genomics}, 44\penalty0 (19):\penalty0 915--924,
  2012.

\bibitem[Smith(2006)]{smith2006network}
Reginald~D Smith.
\newblock The network of collaboration among rappers and its community
  structure.
\newblock \emph{Journal of Statistical Mechanics: Theory and Experiment},
  2006\penalty0 (02):\penalty0 P02006, 2006.

\bibitem[Radrich et~al.(2010)Radrich, Tsuruoka, Dobson, Gevorgyan, Swainston,
  Baart, and Schwartz]{radrich2010integration}
Karin Radrich, Yoshimasa Tsuruoka, Paul Dobson, Albert Gevorgyan, Neil
  Swainston, Gino Baart, and Jean-Marc Schwartz.
\newblock Integration of metabolic databases for the reconstruction of
  genome-scale metabolic networks.
\newblock \emph{BMC systems biology}, 4\penalty0 (1):\penalty0 114, 2010.

\bibitem[Schott(2007)]{schott2007test}
James~R Schott.
\newblock A test for the equality of covariance matrices when the dimension is
  large relative to the sample sizes.
\newblock \emph{Computational Statistics \& Data Analysis}, 51\penalty0
  (12):\penalty0 6535--6542, 2007.

\bibitem[Li et~al.(2012)Li, Chen, et~al.]{li2012two}
Jun Li, Song~Xi Chen, et~al.
\newblock Two sample tests for high-dimensional covariance matrices.
\newblock \emph{The Annals of Statistics}, 40\penalty0 (2):\penalty0 908--940,
  2012.

\bibitem[Cai et~al.(2013)Cai, Liu, and Xia]{cai2013two}
Tony Cai, Weidong Liu, and Yin Xia.
\newblock Two-sample covariance matrix testing and support recovery in
  high-dimensional and sparse settings.
\newblock \emph{Journal of the American Statistical Association}, 108\penalty0
  (501):\penalty0 265--277, 2013.

\bibitem[Cai and Liu(2016)]{cai2016large}
T~Tony Cai and Weidong Liu.
\newblock Large-scale multiple testing of correlations.
\newblock \emph{Journal of the American Statistical Association}, 111\penalty0
  (513):\penalty0 229--240, 2016.

\bibitem[Zhu et~al.(2017)Zhu, Lei, Devlin, and Roeder]{zhu2017testing}
Lingxue Zhu, Jing Lei, Bernie Devlin, and Kathryn Roeder.
\newblock Testing high-dimensional covariance matrices, with application to
  detecting schizophrenia risk genes.
\newblock \emph{The Annals of Applied Statistics}, 11\penalty0 (3):\penalty0
  1810, 2017.

\bibitem[Subramanian et~al.(2005)Subramanian, Tamayo, Mootha, Mukherjee, Ebert,
  Gillette, Paulovich, Pomeroy, Golub, Lander, et~al.]{subramanian2005gene}
Aravind Subramanian, Pablo Tamayo, Vamsi~K Mootha, Sayan Mukherjee, Benjamin~L
  Ebert, Michael~A Gillette, Amanda Paulovich, Scott~L Pomeroy, Todd~R Golub,
  Eric~S Lander, et~al.
\newblock Gene set enrichment analysis: a knowledge-based approach for
  interpreting genome-wide expression profiles.
\newblock \emph{Proc Natl Acad Sci}, 102\penalty0 (43):\penalty0 15545--15550,
  2005.

\bibitem[Mootha et~al.(2003)Mootha, Lindgren, Eriksson, Subramanian, Sihag,
  Lehar, Puigserver, Carlsson, Ridderstr{\aa}le, Laurila,
  et~al.]{mootha2003pgc}
Vamsi~K Mootha, Cecilia~M Lindgren, Karl-Fredrik Eriksson, Aravind Subramanian,
  Smita Sihag, Joseph Lehar, Pere Puigserver, Emma Carlsson, Martin
  Ridderstr{\aa}le, Esa Laurila, et~al.
\newblock Pgc-1$\alpha$-responsive genes involved in oxidative phosphorylation
  are coordinately downregulated in human diabetes.
\newblock \emph{Nature genetics}, 34\penalty0 (3):\penalty0 267--273, 2003.

\bibitem[Consortium et~al.(2004)]{gene2004gene}
Gene~Ontology Consortium et~al.
\newblock The gene ontology (go) database and informatics resource.
\newblock \emph{Nucleic acids research}, 32\penalty0 (suppl 1):\penalty0
  D258--D261, 2004.

\bibitem[Kanehisa and Goto(2000)]{kanehisa2000kegg}
Minoru Kanehisa and Susumu Goto.
\newblock Kegg: kyoto encyclopedia of genes and genomes.
\newblock \emph{Nucleic acids research}, 28\penalty0 (1):\penalty0 27--30,
  2000.

\bibitem[Khatri et~al.(2012)Khatri, Sirota, and Butte]{khatri2012ten}
Purvesh Khatri, Marina Sirota, and Atul~J Butte.
\newblock Ten years of pathway analysis: current approaches and outstanding
  challenges.
\newblock \emph{PLoS computational biology}, 8\penalty0 (2):\penalty0 e1002375,
  2012.

\bibitem[Meinshausen and B{\"u}hlmann(2006)]{meinshausen2006high}
Nicolai Meinshausen and Peter B{\"u}hlmann.
\newblock High-dimensional graphs and variable selection with the lasso.
\newblock \emph{The Annals of Statistics}, pages 1436--1462, 2006.

\bibitem[Friedman et~al.(2008)Friedman, Hastie, and
  Tibshirani]{friedman2008sparse}
Jerome Friedman, Trevor Hastie, and Robert Tibshirani.
\newblock Sparse inverse covariance estimation with the graphical lasso.
\newblock \emph{Biostatistics}, 9\penalty0 (3):\penalty0 432--441, 2008.

\bibitem[Yuan and Lin(2007)]{yuan2007model}
Ming Yuan and Yi~Lin.
\newblock Model selection and estimation in the gaussian graphical model.
\newblock \emph{Biometrika}, 94\penalty0 (1):\penalty0 19--35, 2007.

\bibitem[Banerjee et~al.(2008)Banerjee, Ghaoui, and
  d’Aspremont]{banerjee2008model}
Onureena Banerjee, Laurent~El Ghaoui, and Alexandre d’Aspremont.
\newblock Model selection through sparse maximum likelihood estimation for
  multivariate gaussian or binary data.
\newblock \emph{Journal of Machine learning research}, 9\penalty0
  (Mar):\penalty0 485--516, 2008.

\bibitem[Rothman et~al.(2008)Rothman, Bickel, Levina, Zhu,
  et~al.]{rothman2008sparse}
Adam~J Rothman, Peter~J Bickel, Elizaveta Levina, Ji~Zhu, et~al.
\newblock Sparse permutation invariant covariance estimation.
\newblock \emph{Electronic Journal of Statistics}, 2:\penalty0 494--515, 2008.

\bibitem[Levina et~al.(2008)Levina, Rothman, and Zhu]{levina2008sparse}
Elizaveta Levina, Adam Rothman, and Ji~Zhu.
\newblock Sparse estimation of large covariance matrices via a nested lasso
  penalty.
\newblock \emph{The Annals of Applied Statistics}, pages 245--263, 2008.

\bibitem[NCI and NHGRI(2018)]{TCGA}
NCI and NHGRI.
\newblock The cancer genome atlas, 2018.
\newblock URL \url{https://cancergenome.nih.gov}.
\newblock National Cancer Institute and National Human Genome Research
  Institute, The Cancer Genome Atlas.

\bibitem[Frankl and Maehara(1990)]{Frankl1990}
Peter Frankl and Hiroshi Maehara.
\newblock Some geometric applications of the beta distribution.
\newblock \emph{Annals of the Institute of Statistical Mathematics},
  42:\penalty0 463--474, 1990.

\bibitem[Yeung et~al.(2002)Yeung, Tegn{\'e}r, and Collins]{yeung2002reverse}
MK~Stephen Yeung, Jesper Tegn{\'e}r, and James~J Collins.
\newblock Reverse engineering gene networks using singular value decomposition
  and robust regression.
\newblock \emph{Proceedings of the National Academy of Sciences}, 99\penalty0
  (9):\penalty0 6163--6168, 2002.

\bibitem[Bar and Schifano(2019)]{BarSchifano}
Haim Bar and Elizabeth~D. Schifano.
\newblock Differential variation and expression analysis.
\newblock \emph{Stat}, 8\penalty0 (1):\penalty0 e237, 2019.
\newblock \doi{10.1002/sta4.237}.
\newblock URL \url{https://onlinelibrary.wiley.com/doi/abs/10.1002/sta4.237}.

\bibitem[Dempster et~al.(1977)Dempster, Laird, and Rubin]{EM}
A.~P. Dempster, N.~M. Laird, and D.~B. Rubin.
\newblock {Maximum likelihood from incomplete data via the EM algorithm}.
\newblock \emph{Journal of the Royal Statistical Society, Series B},
  39\penalty0 (1):\penalty0 1--38, 1977.

\bibitem[Benjamini and Hochberg(1995)]{benjamini:hochberg}
Yoav Benjamini and Yosef Hochberg.
\newblock Controlling the false dicovery rate-a practical and powerful approach
  to multiple testing.
\newblock \emph{Journal of the Royal Statistical Society Series B}, 57\penalty0
  (3):\penalty0 499--517, 1995.

\bibitem[Zhao et~al.(2015)Zhao, Li, Liu, Roeder, Lafferty, and
  Wasserman]{rpackagehuge}
Tuo Zhao, Xingguo Li, Han Liu, Kathryn Roeder, John Lafferty, and Larry
  Wasserman.
\newblock \emph{huge: High-Dimensional Undirected Graph Estimation}, 2015.
\newblock URL \url{https://CRAN.R-project.org/package=huge}.
\newblock R package version 1.2.7.

\bibitem[Barab{\'a}si and Albert(1999)]{barabasi1999emergence}
Albert-L{\'a}szl{\'o} Barab{\'a}si and R{\'e}ka Albert.
\newblock Emergence of scaling in random networks.
\newblock \emph{Science}, 286\penalty0 (5439):\penalty0 509--512, 1999.

\bibitem[Allahyar et~al.(2019)Allahyar, Ubels, and de~Ridder]{Allahyar2019}
Amin Allahyar, Joske Ubels, and Jeroen de~Ridder.
\newblock A data-driven interactome of synergistic genes improves network-based
  cancer outcome prediction.
\newblock \emph{PLOS Computational Biology}, 15\penalty0 (2):\penalty0 1--21,
  02 2019.
\newblock \doi{10.1371/journal.pcbi.1006657}.
\newblock URL \url{https://doi.org/10.1371/journal.pcbi.1006657}.

\bibitem[Staiger et~al.(2013)Staiger, Cadot, Györffy, Wessels, and Klau]{ACES}
Christine Staiger, Sidney Cadot, Balázs Györffy, Lodewyk Wessels, and Gunnar
  Klau.
\newblock Current composite-feature classification methods do not outperform
  simple single-genes classifiers in breast cancer prognosis.
\newblock \emph{Frontiers in Genetics}, 4:\penalty0 289, 2013.
\newblock ISSN 1664-8021.
\newblock \doi{10.3389/fgene.2013.00289}.
\newblock URL
  \url{https://www.frontiersin.org/article/10.3389/fgene.2013.00289}.

\bibitem[Towles et~al.(2016)Towles, Clark, Wahlin, Uttamsingh, Rettie, and
  Jackson]{towles2016cytochrome}
Joanna~K Towles, Rebecca~N Clark, Michelle~D Wahlin, Vinita Uttamsingh, Allan~E
  Rettie, and Klarissa~D Jackson.
\newblock Cytochrome p450 3a4 and cyp3a5-catalyzed bioactivation of lapatinib.
\newblock \emph{Drug Metabolism and Disposition}, 44\penalty0 (10):\penalty0
  1584--1597, 2016.

\bibitem[Huszno et~al.(2015)Huszno, Badora, and Nowara]{huszno2015influence}
Joanna Huszno, Agnieszka Badora, and El{\.z}bieta Nowara.
\newblock The influence of steroid receptor status on the cardiotoxicity risk
  in her2-positive breast cancer patients receiving trastuzumab.
\newblock \emph{Archives of medical science: AMS}, 11\penalty0 (2):\penalty0
  371, 2015.

\bibitem[Fisher et~al.(2010)Fisher, Ultsch, Lingel, Schaefer, Shao, Birtalan,
  Sidhu, and Eigenbrot]{fisher2010structure}
Robert~D Fisher, Mark Ultsch, Andreas Lingel, Gabriele Schaefer, Lily Shao,
  Sara Birtalan, Sachdev~S Sidhu, and Charles Eigenbrot.
\newblock Structure of the complex between her2 and an antibody paratope formed
  by side chains from tryptophan and serine.
\newblock \emph{Journal of molecular biology}, 402\penalty0 (1):\penalty0
  217--229, 2010.

\bibitem[Miolo et~al.(2016)Miolo, Muraro, Caruso, Crivellari, Ash, Scalone,
  Lombardi, Rizzolio, Giordano, and Corona]{miolo2016pharmacometabolomics}
Gianmaria Miolo, Elena Muraro, Donatella Caruso, Diana Crivellari, Anthony Ash,
  Simona Scalone, Davide Lombardi, Flavio Rizzolio, Antonio Giordano, and
  Giuseppe Corona.
\newblock Pharmacometabolomics study identifies circulating spermidine and
  tryptophan as potential biomarkers associated with the complete pathological
  response to trastuzumab-paclitaxel neoadjuvant therapy in her-2 positive
  breast cancer.
\newblock \emph{Oncotarget}, 7\penalty0 (26):\penalty0 39809, 2016.

\bibitem[Cook et~al.(2011)Cook, Shajahan, and Clarke]{cook2011autophagy}
Katherine~L Cook, Ayesha~N Shajahan, and Robert Clarke.
\newblock Autophagy and endocrine resistance in breast cancer.
\newblock \emph{Expert review of anticancer therapy}, 11\penalty0 (8):\penalty0
  1283--1294, 2011.

\bibitem[Anurag et~al.(2018)Anurag, Punturi, Hoog, Bainbridge, Ellis, and
  Haricharan]{anurag2018comprehensive}
Meenakshi Anurag, Nindo Punturi, Jeremy Hoog, Matthew~N Bainbridge, Matthew~J
  Ellis, and Svasti Haricharan.
\newblock Comprehensive profiling of dna repair defects in breast cancer
  identifies a novel class of endocrine therapy resistance drivers.
\newblock \emph{Clinical Cancer Research}, 24\penalty0 (19):\penalty0
  4887--4899, 2018.

\end{thebibliography}

	\beginsupplement
    \section*{Supplementary Materials}
    Web-based Supplementary Materials for `A Mixture Model to Detect Edges in Sparse Co-expression Graphs' by Haim Bar and Seojin Bang.

    \begin{figure}[h]
	\centering
	\includegraphics[trim={0 0.6cm 0 1.2cm},clip,scale=0.8]{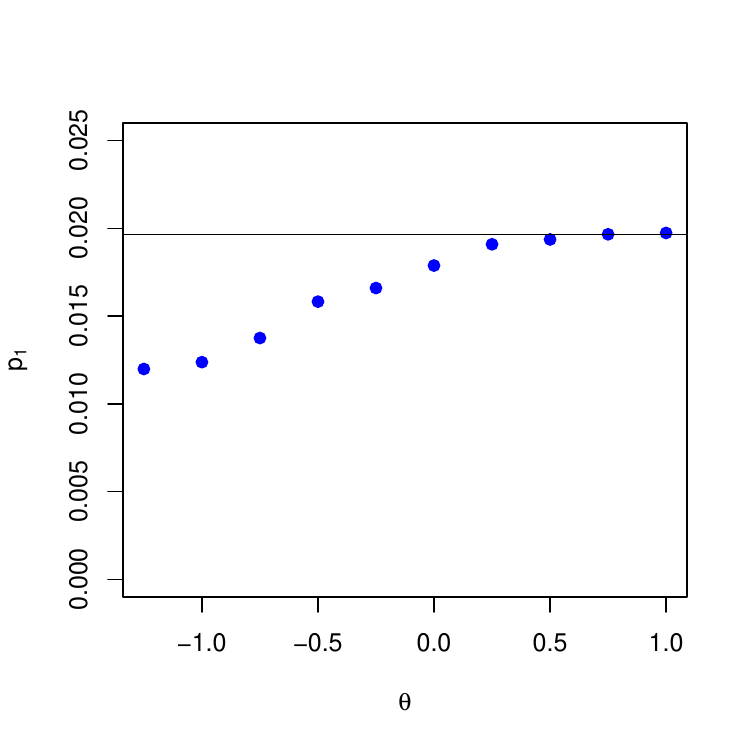}
	\caption{The average
estimate of $p_1$, the proportion of positively correlated pairs of genes, from 20 replications, is plotted versus $\theta$ in the \textit{two negatively correlated blocks}
configuration. The horizontal black line depicts the true value of $p_1$.}\label{fig.fittedp1}
\end{figure}
    \newpage
    	
	\newpage
	
	\begin{figure}[h]
	\centering
	\includegraphics[width=0.9\textwidth]{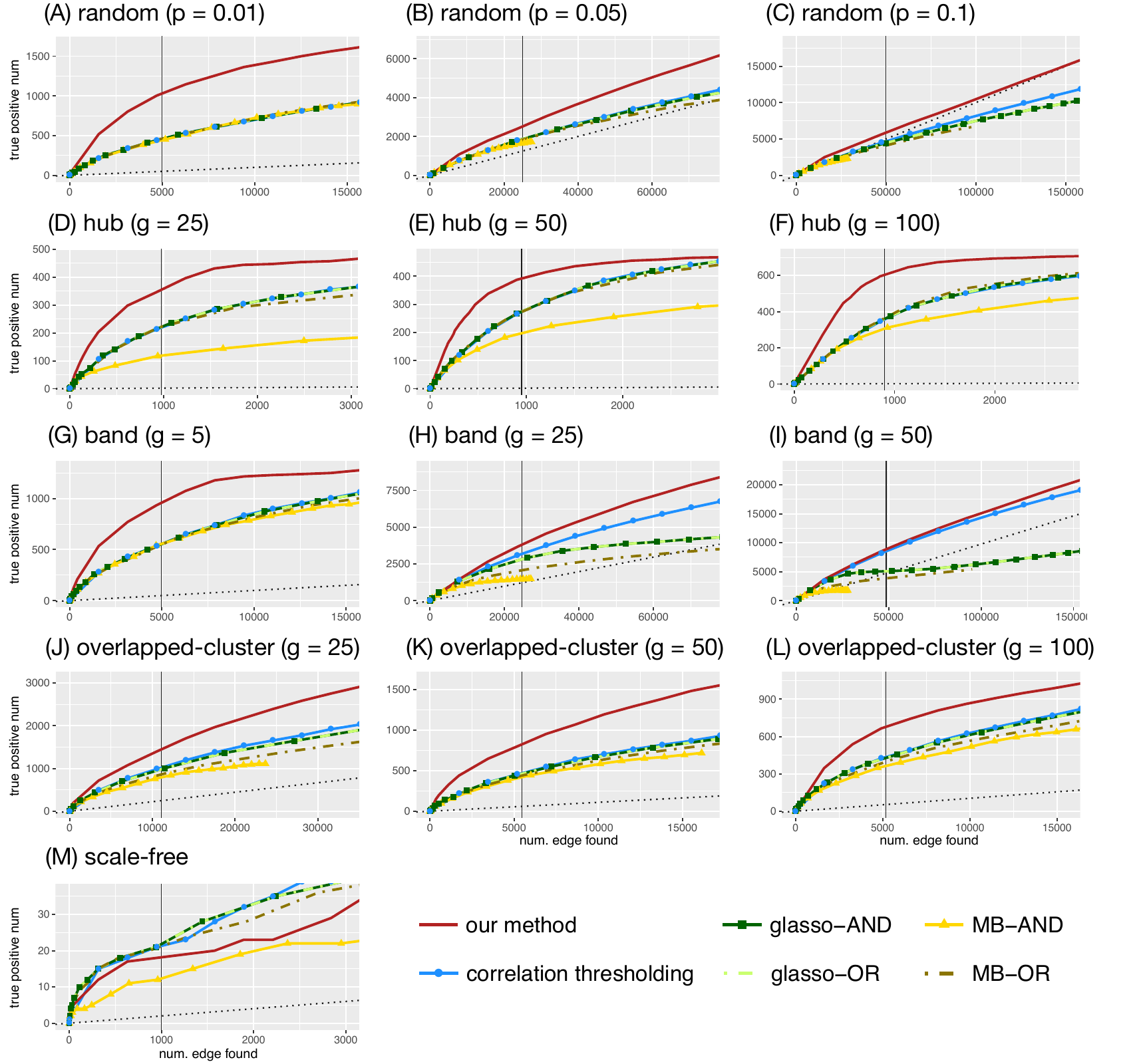}
	\caption{The numbers of true positive edges given the total number of edges identified by each method. The true matrix is obtained by applying a threshold to the true covariance matrix. The y-axis represents the number of true positive edges and the x-axis represents the total number of edges identified. The vertical line represents the number of true edges. The black dotted line is a regression line with 0 intercept  and  slope equal to the true sparsity, which represents the expected number of true positive edges when the edges are identified in a random manner.}\label{fig.sim.TPcovmat}
\end{figure}

	\newpage

	%
	
	\newpage
	
    \begin{figure}[h]
	\centering
	\includegraphics[trim={0cm 0cm 0cm 0cm},clip,scale=1]{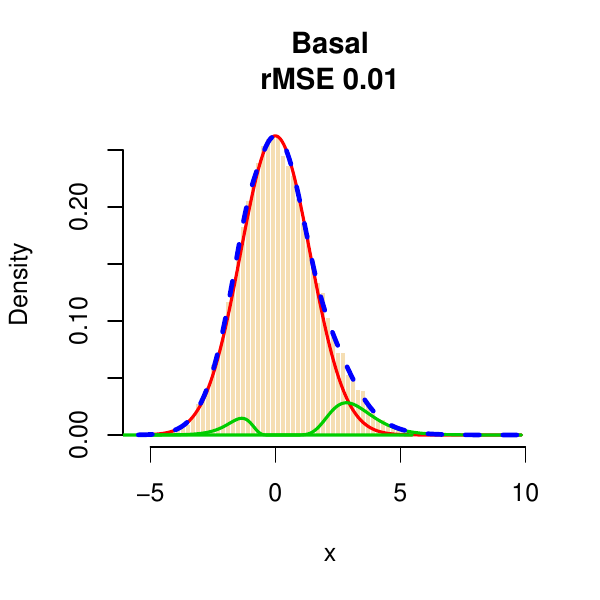}
	\caption{Goodness of fit plot of $L_2N$ mixture model. The distributions of $w_{mn}=arctanh(r_{mn})$ for the Basal group. The red curve represents the null component, the green curves represent the nonull components, and the dashed blue line represents the fitted mixture distribution.}\label{fig.fittedmixture}
\end{figure}
    \newpage
    
	\begin{figure}[h]
	\centering
	\includegraphics[trim={0cm 0cm 0cm 0cm},clip,scale=1]{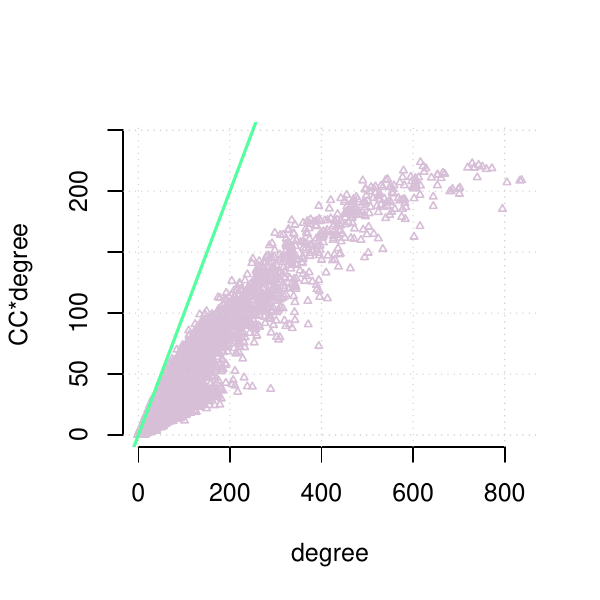}
	\caption{Plot of $\gamma_m d_m$ versus $d_m$ for the Luminal A group.}
	\end{figure}

		\begin{figure}[h]
	\centering
	\includegraphics[trim={0cm 0cm 0cm 0cm},clip,scale=1]{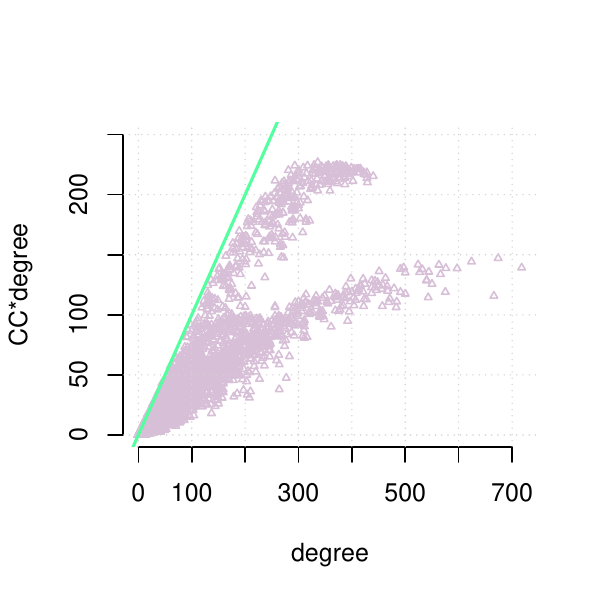}
	\caption{Plot of $\gamma_m d_m$ versus $d_m$ for the Luminal B group.}
	\end{figure}

	\begin{figure}[h]
	\centering
	\includegraphics[trim={0cm 0cm 0cm 0cm},clip,scale=1]{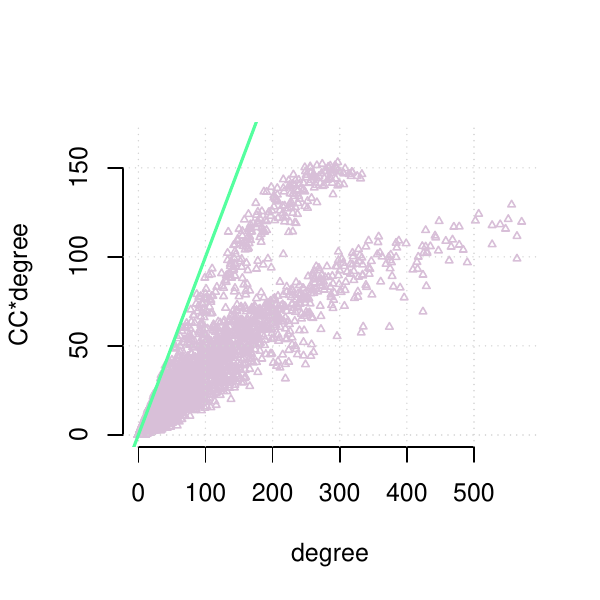}
	\caption{Plot of $\gamma_m d_m$ versus $d_m$ for the Basal group.}
	\end{figure}
	\newpage
	
    \begin{table}[ht]
	\centering
	\caption[caption]{Pathways characterizing the normal group.} \label{table:pathwaylistnormal}
	\begin{adjustbox}{max width=\textwidth}
		\def\arraystretch{0.95}
		\begin{threeparttable}
			\begin{tabular}{@{}lcccp{13cm}@{}}
			\toprule
			Pathway Name & Pathway Size & Cluster Index & P-value  & Genes \cr
			\midrule
		Endocytosis                               & 181          & 35            & 0.001379 & HLA-B, HLA-C, HLA-E, HLA-F, HLA-G \cr
		Lysosome                                  & 109          & 20            & 0.007482 & CTSB, CTSL, MAN2B1, LGMN \cr
		Primary immunodeficiency                  
		& 34        & 4             & 6.69E-08 & ADA, BTK, CD3D, CD4, CD8A, CD40, IL2RG, LCK, PTPRC \cr
		&  			& 14            & 1.81E-06 & CD8B, CD79A, IL7R, ZAP70, ICOS \cr
		T cell receptor signaling pathway
		& 103       & 4             & 1.34E-06 & CD3D, CD4, CD8A, CD28, FYN, ITK, LCK, LCP2, PIK3CD, PTPN6, PTPRC, VAV1 \cr
		&           & 14            & 0.000249 & CD247, CD8B, IFNG, ZAP70, ICOS \cr	
		Graft-versus-host disease  
		& 36        & 4             & 9.07E-08 & CD28, CD86, HLA-A, HLA-DMA, HLA-DMB, HLA-DPA1, HLA-DPB1, HLA-DRA, HLA-DRB1 \cr               
		&           & 14            & 0.001665 & IFNG, KLRD1, PRF1 \cr	
		Natural killer cell mediated cytotoxicity 
		& 119       & 4             & 3.20E-08 & CD48, FCER1G, FYN, HLA-A, ITGAL, ITGB2, LCK, LCP2, SH2D1A, PIK3CD, PRKCB, PTPN6, RAC2, TYROBP, VAV1 \cr
		&           & 14            & 0.000336 & CD247, IFNG, KLRD1, PRF1, ZAP70 \cr
		&           & 35            & 0.002522 & HLA-B, HLA-C, HLA-E, HLA-G \cr
		Antigen processing and presentation       
		& 69        & 35            & 1.61E-09 & B2M, HLA-B, HLA-C, HLA-E, HLA-F, HLA-G, PSME2, TAP1 \cr
		&           & 4             & 1.80E-07 & CD4, CD8A, CD74, CTSS, HLA-A, HLA-DMA, HLA-DMB, HLA-DPA1, HLA-DPB1, HLA-DRA, HLA-DRB1 \cr
		&           & 20            & 0.001641 & CTSB, CTSL, LGMN, IFI30 \cr
		&           & 14            & 0.007611 & CD8B, IFNG, KLRD1 \cr
		Phagosome  
		& 138       & 4             & 4.67E-06 & CTSS, FCGR2B, HLA-A, HLA-DMA, HLA-DMB, HLA-DPA1, HLA-DPB1, HLA-DRA, HLA-DRB1, ITGB2, NCF4, CORO1A, CLEC7A\cr		     
		&        	& 35            & 3.40E-05 & HLA-B, HLA-C, HLA-E, HLA-F, HLA-G, TAP1 \cr 
		&           & 20            & 0.000366 & ATP6V1B2, CTSL, FCGR2A, FCGR3B, MSR1, NCF2 \cr                      
		&           & 11            & 0.003818 & C1R, COMP, ITGA5, ITGAV, THBS2, MRC2 \cr\bottomrule  
		\end{tabular}
		\begin{tablenotes}
			\footnotesize
			\item Uniquely enriched pathways (\textit{endocytosis} and \textit{lysosome}) and enriched in a larger number of clusters in the normal group than other subtypes (\textit{phagosome}, \textit{antigen processing and presentation}, \textit{natural killer cell mediated cytotoxicity}, \textit{graft-versus-host disease}, \textit{T cell receptor signaling pathway}, and \textit{primary immunodeficiency}). All identified clusters are indexed by their size. The cluster index indicates the order of the clusters. Note that a pathway can be enriched in multiple clusters as we performed the enrichment analysis for each cluster. P-values are adjusted to control the false discovery rate using \citet{benjamini:hochberg}.
		\end{tablenotes}
		\end{threeparttable}
	\end{adjustbox}	
\end{table}

    \newpage
    


\end{document}